# Quantum Light in Curved Low Dimensional Hexagonal Boron Nitride Systems

*Nathan Chejanovsky [1,2], Youngwook Kim [2], Andrea Zappe [1], Benjamin Stuhlhofer [2], Takashi Taniguchi [3], Kenji Watanabe [3], Durga Dasari [1,2], Amit Finkler\* [1], Jurgen H. Smet [2] and Jörg Wrachtrup [1,2]*

[1] 3. Physikalisches Institut, Universität Stuttgart, Pfaffenwaldring 57, 70569 Stuttgart, Germany

[2] Max Planck Institute for Solid State Research, Heisenbergstr. 1, 70569 Stuttgart, Germany

[3] National Institute for Materials Science, 1-1 Namiki, Tsukuba, 305-0044, Japan

**Abstract**

Low-dimensional wide bandgap semiconductors open a new playing field in quantum optics using sub-bandgap excitation. In this field, hexagonal boron nitride (h-BN) has been reported to host single quantum emitters (QEs), linking QE density to perimeters. Furthermore, curvature/perimeters in transition metal dichalcogenides (TMDCs) have demonstrated a key role in QE formation. We investigate a curvature-abundant BN system – quasi one-dimensional BN nanotubes (BNNTs) fabricated via a catalyst-free method. We find that non-treated BNNT is an abundant source of stable QEs and analyze their emission features down to single nanotubes, comparing dispersed/suspended material. Combining high spatial resolution of a scanning electron microscope, we categorize and pin-point emission origin to a scale of less than 20 nm, giving us a one-to-one validation of emission source with dimensions smaller than the laser excitation wavelength, elucidating nano-antenna effects. Two emission origins emerge: hybrid/entwined BNNT. By artificially curving h-BN flakes, similar QE spectral features are observed. The impact on emission of solvents used in commercial products and curved regions is also demonstrated. The 'out of the box' availability of QEs in BNNT, lacking processing contamination, is a milestone for unraveling their atomic features. These findings open possibilities for precision engineering of QEs, puts h-BN under a similar 'umbrella' of TMDC's QEs and provides a model explaining QEs spatial localization/formation using electron/ion irradiation and chemical etching.

**Introduction**

BNNTs [1] can be seen as the 2D h-BN hexagonal grid rolled into a closed nanotube structure, [2] making them a natural candidate to explore the effect of curvature in BN systems. The 1D nature of BNNTs means that the system should be perimeter (edges, boundaries) - [3] as well as curvature-abundant. Both have in-plane $sp^2$ orbitals with similar bond lengths: BNNT - 1.44 Å and h-BN - 1.45 Å. [4] With increasing diameter, the formation energy of BNNTs approaches that of 2D h-BN. [5] BNNTs are also analogues to the more explored system of carbon nanotubes (shown to harbor QEs) [6] with the distinct difference of a wide bandgap of ~ 5 eV, which is independent of non-deformed nanotube geometry [7] and helicity. [2,8] BNNTs broken sub-lattice symmetry gives rise to a macroscopic electric polarization, whose ground state polarization is an intrinsically nonlocal quantum effect. [9] Small diameter BNNTs buckle - B atoms move inward and N atoms outward - resulting in a dipolar double cylinder shell of a negative/positive outer/inner N/B cylinder, respectively. [4,8] In an analogous manner, for monolayer h-BN, a dipolar electric distribution has been shown to occur above and below the monolayer, probed by nuclear quadrupole resonance (nuclear spin ≥ 1 in the lattice) using a nitrogen vacancy color center in diamond. [10] BNNTs have shown intriguing physics such as the giant Stark effect [11] and transformation of BNNTs from insulating to semiconducting using physical deformation.[7] Due to their high temperature stability, light weight, high stress endurance, resistance to oxidation at non-defective sites, [12] biocompatibility and potential use as neutron shields/detectors, BNNTs are ideal for emerging space technologies [13] as well as for biological cell research. [14] Photo-luminescence of BN nano-whiskers, using cathodoluminescence, has shown emission in the visible range attributed to point defects, quantum confinement and $sp^3$ bonding. [15] For BNNTs, exciton emission was shown for deep ultraviolet wavelengths. [16] Using near bandgap excitation, defect bound excitons' emission was independent of nanotube diameter and wall thickness (for at least 20 layers) and optically shown to behave as pieces of curved 2D h-BN. [17] Above bandgap excitation energy has shown hints of $sp^3$ bonded defects contributing to emission, [18] whereas X-ray excited optical luminescence was also sensitive to tube curvature. [19] Oxygen defect related radiative transitions have also been proposed. [20,21,22] BNNT transmission electron microscopy (TEM) research has visualized induced elastic deformation in tubes for angles larger than 30 degrees, [23] point defects such as boron and nitrogen vacancies, BN di-vacancies (reconstructing to Stone-Wales [24] structures), [25,3] pentagons and heptagons atomic configurations (Stone-Wales) [24,3] at tube terminations, [26,27] and a range of helices (from zig-zag to arm-chair) within a single tube. [28] Less destructive spectroscopy methods combining AFM and near-field infrared scattering have also shown the presence of twists and structural defects. [5] From these atomic resolution and optical observations, it is clear that point like atomic defect species formation in this flexible nano-material is intimately connected to geometry, curvature, strain and proximity to the environment due to its large surface area. Recently, it has also been shown that BNNT can serve as a scaffold for other colloidal wide bandgap semiconductors such as ZnO, [29] thus creating hybrid BNNT. Nevertheless, BNNT research is relatively in its primary stages, due to only recently emerging reliable fabrication methods and difficulties arising in separating the nanotubes after fabrication, [30] which are clustered in a semi 'cotton' like fashion in the bulk. Here we make use of commercially available BNNTs fabricated using a catalyst-free high temperature pressure (HTP) laser heating method [26] and use them to further study single QEs [31,32,3,33,34,35,36,37,38,39,40,41,42] in the BN hexagonal systems.

The manuscript is divided as follows: BNNT properties are explained followed by a detailed description of the role oxygen has in BN systems. Then, spectral analysis of emission properties of bulk and micro-bundles of BNNT QEs is shown. Afterwards we show that by dispersion methods, quantum emission can be categorized in two PL spectral classes: one with relatively narrow features and the other with broader features. For each class, we conduct a spectral, phonon and lifetime analysis. These are supported by SEM imaging, confirming our classifications. We note that previous reports have shown electron irradiation to induce SQEs in h-BN [34,40], therefore all SEM imaging performed here were done after PL measurements. Finally, we demonstrate the role of curvature in low dimensional BN systems by artificially curving 2D h-BN flakes on diamond nanopillars and a $ZrO_2$ hemisphere. All measurements were done at ambient conditions using sub-bandgap excitation energies.

*BNNT structure and components*

HTP-fabricated BNNTs come in a 'cotton' like form, consisting of numerous entwined tubes (see SI for an example). We define a 'cotton' ball of as received material with a radius of ~ > 5 cm as 'bulk' BNNT. HTP fabrication is advantageous since it is catalyst-free, produces high quality BNNTs, and minimizes defects from foreign atomic species (such as carbon) [26,43] to the BN lattice as opposed to other methods (See Ref. 13 for further details). Typical wall thicknesses range between 2 to 5 layers, with diameters of 3 nm to 6 nm and interspacing wall distances of 3.4 Å. [26] Single nanotubes can be up to 200 µm long. Nevertheless, the material also contains 2-dimensional (2D) h-BN, typically in the lateral size range of 50 nm to 200 nm and also boron nanoparticles, visualized by TEM for HTP BNNT, [2,13,44,45] as well as for various fabrication techniques. [2] In addition, for annealed HTP BNNT, 2D BN oxide platelets have been visualized. [12] For simplicity we denote these nanomaterials, attached to the BNNT tubular structure, as **hybrid** material in the manuscript and discuss their implication. For the HTP fabrication method the measured bandgap is 5.74 eV, [13] similar to that of 2D h-BN (5.95 eV). [46]

### *The role of oxygen in BN hexagonal systems*

Photoluminescence (PL) comparative studies in vacuum and ambient conditions on TMDCs [47] and h-BN [48] have shown that PL is strongly affected by ambient gas molecules, attributed to exterior surface defects exposed to reactive species which physisorb, such as $O_2$. An oxygen healing mechanism in h-BN for nitrogen vacancies (oxygen substituting nitrogen vacancies) has been observed in PL [48] studies and TEM microscopy, [49,50] with a time scale of 0.6 µs once exposed to ambient conditions, whereas h-BN was indifferent to $N_2$. [48] In h-BN, stabilizing SQEs has been attributed to high temperature annealing (usually 850 °C). Intuitively, this annealing should remove oxygen, however this procedure is done in an inert gas environment (typically argon), [33,34,3,35,37,38,40,41,42] and not in vacuum. The only report to of annealing in vacuum, resulting in less stable SQEs is described in Ref. 3. Studies on annealing of $WS_2$ in an inert argon environment have shown SQE induced on *perimeters*, attributed to formation of $WO_3$, due to residual water/oxygen in the annealing chamber. [51] It has been suggested there that similar SQEs species could be found in $WO_3$ crystals. *Perimeter* oxidation has been shown to be preferential in an independent study on $WS_2$ and $WSe_2$. [52]

Core level spectroscopy reveals the effects of oxidizing environments on 2D h-BN in oxygen, demonstrating the creation of $BN_2O$, $B-NO_2$ and $B-O_3$ species as a substituting mechanism for nitrogen sites. A post-anneal in vacuum at 600 °C demonstrated that $B_2O_3$ is formed and oxygen is not removed with a complete destruction of the monolayer. [53] Using x-ray absorption, sputtering h-BN in an inert environment resulted also in $B_2O_3$ structures, attributed also to residual water/$O_2$ in the chamber, whereas using a prolonged pre-bake of the preparation chamber reduced the observed oxygen. [54] To the best of our knowledge, in all the annealing procedures to date of h-BN SQEs, no long pre-bake of the annealing chamber was performed prior to annealing in inert gas. [33,34,3,35,37,38,40,41,42] Different studies of annealing in air revealed that oxidation can occur for temperatures above 700 °C for h-BN [55] and BNNT, forming the aforementioned 2D hybrid oxide platelets. [12] Furthermore, a PL study for SQEs in h-BN, post-annealed in vacuum (after annealing at 850 °C in argon) has shown stable SQEs up to (526 °C) 800 K, [41] hinting to a connection of the observed temperature stability of oxygen in the other study. [53] It is has even been shown that SQE creation is correlated with increasing annealing temperatures in argon with a maximum yield at temperatures of 1000 °C – well above 700 °C. [34] We note that in these oxidation studies [53,54,55] the threshold temperature for oxidation is given from data acquired from numerous oxygen defects and does not reveal the threshold for creation of point-like minute nano-$B_2O_3$. The SQE studies [33,34,3,35,37,38,40,41,42] suggest that although oxygen was not directly used in annealing procedures, minute traces of oxygen were present, possibly creating nano-$B_2O_3$ structures. In addition, X-ray absorption ion irradiation studies on h-BN [56] and BNNT [57], have demonstrated oxygen healing once re-exposed to ambient conditions with irradiation favorably creating nitrogen vacancies. [56] h-BN irradiation was shown to also induce SQEs in other studies. [3,40,42] In these SQE studies, irradiating and annealing were independent, [3,40,42] therefore exposure to ambient conditions exceeded a time scale of 0.6 μs [48] prior to annealing, allowing oxygen healing of freshly induced defective sites.

In our BNNT material, composed of curved h-BN units and also 2D h-BN, [2,13,44,45] the surface area is higher than a bulk (or exfoliated) h-BN crystal. The concentration of exterior surface defects is therefore also higher and healed by ambient oxygen. The high temperature method used to create BNNT, annealing done in previous SQE studies of 2D h-BN [33,34,3,35,37,38,40,41,42] (as well as in this study, see below and SI) to stabilize emitters, [48] chemical oxidation etching shown to induce SQEs, [3] oxygen related photochemistry [31] of photo-induced modifications of SQEs to blue wavelengths, [37] in conjunction with observations written above, warrant an optical investigation of $B_2O_3$ to evaluate the role oxygen can play in semiconductors of the boron family. Oxygen can be introduced during fabrication, exposure to air prior to annealing, during annealing in chambers with traces of water/$O_2$ or during the processing of h-BN in solvents/liquids (processing sensitivity is demonstrated in the SI). The most common form of $B_2O_3$ is vitreous (v-$B_2O_3$) [58,59], since spontaneous crystallization of $B_2O_3$ is not a simple process. [60] v-$B_2O_3$ contains $BO_3$ and planar boroxol rings (an analogous h-BN ring with O substituting N in the center of $B_3O_6$, Schematic – Fig. 1a), with a bandgap of ~ 6 eV. [60] These planar rings can be seen as disordered planar 2D material within the bulk.

## Results

*Bulk BNNT*

As a first step, bulk BNNT (see Methods) was probed using 735 nm Raman excitation. We confirm the Raman-Stokes shift of BNNT with the main peak at 1379 cm$^{-1}$ (Fig. 1a) corresponding to the E$_{2g}$ in-plane Raman modes, with a similar PL structure to those seen in previous infrared Raman studies. [44,61,62] We surmise that some sp$^3$ hybridization is present and the peak at ~ 3200 cm$^{-1}$ can correspond to B-OH/B-O bonds, [44,63] or possibly to moisture. [12,44] For clarification, on the wavenumber axis in Fig. 1a we show the known Raman shifts and optical phonons for four species of the boron family: h-BN, BNNT, cubic BN (c-BN) and Boric acid (c-BN is an sp$^3$ allotrope of BN which is isoelectric to diamond [64,65]). See SI for a detailed analysis of this PL. An energy-dispersive X-ray spectroscopy (EDX) measurement on bulk BNNT (Fig. 1a – inset) identifies 3 major atomic components: B/N (0.17/0.39 eV, respectively), as expected, and oxygen (0.53 eV), in agreement with our Raman measurements. Consistent with previous studies, [26,43] the carbon signal (0.27 eV) is extremely weak. Upon 532 nm CW excitation, non-bleaching emitters are immediately evident, with no slow-rate bleaching. We observe two typical PL spectral features, which as we will show, originate from **hybrid material** (narrow PL features) and **entwined BNNTs** (broader PL features). Previous research in 2D h-BN [40] has also suggested that narrow/broad PL features can be an indicator for different emitter classes. However, in the bulk only entwined BNNTs spectra could be discriminated (spectral features are discussed below). We estimate an ensemble QE density using 532 nm excitation of 0.47 μm$^{-2}$ for bulk BNNT. Nevertheless, in the bulk, single QEs could not be isolated.

*Isolating single emitters in BNNT*

Due to these observations, we reduce and disperse the nanotube material. Pulling a small string-like amount of BNNTs, we isolate bundles of nanotubes, whose diameters are on the micron scale. Taking advantage of the electrostatic polar nature of BNNTs, we suspend the micro-bundles on two aluminum needles, thus avoiding substrate contributions while preventing contact with the suspended region. A schematic/microscope image be seen in Fig. 1b, top to bottom, respectively (See SI for PL confocal scan). Remarkably, with the reduction of the amount of BNNT material and the decrease in excitation energy to the wavelength of 594 nm, single QEs (SQE) were isolated. In Fig. 1c-e/f (curve 0)/g we present features of SQEs originating **only from BNNT hybrid material**. Suspended material data is displayed in Figs. 1c-e. An autocorrelation measurement reveals a $g^2(0)$ value of $0.33 \pm 0.01$ (< 0.5) (Fig. 1e – orange curve) indicating the single nature of the emitter. The PL-spectra (Fig. 1c – orange curve) reveals a first sharp peak accompanied with a broader peak. We fit both with Lorentzians (blue and red fillings, respectively). Due to the high coefficient of determination ($R^2 = 0.98$) of the fitting and the similarity to PL features established in previous 2D h-BN research, [33,34,3,35,37,38,40,41,42] we tentatively denote the lowest energetic emission peak as the zero phonon line (ZPL) and the second peak as the phonon side band (PSB). The ZPL is at the wavelength of 639 nm and the PSB is detuned ~ 120 meV (967 cm$^{-1}$) from the ZPL. We note that for some SQEs the entire PSB was absorbed inside the ZPL and could not be distinguished (see SI for example). We estimate a hybrid material SQE density of 0.17 μm$^{-2}$ using 594 nm excitation in micro tubes-bundle.

An analysis of the power-dependent photo-dynamics for the suspended SQE whose PL spectra is displayed in Fig. 1c (orange curve) was done in order to gain insight to whether these are comparable to previous reports in 2D h-BN [3,34,40] See SI for details. We denote our level scheme as 1/2/3 for the ground/excited/metastable states, respectively. $R_{ij}^0$ is the transition between $i \rightarrow j$ for zero excitation power. By analyzing the power dependence of the short/long time scale autocorrelation measurements ($|t| < \sim 30$ ns, $> \sim 30$ ns) and converting them to transition rates, the excited ($\lambda_1$)/metastable ($\lambda_2$) state rate can be extracted, respectively (see Ref. 3 for further details). We get an excited state relaxation rate of $R_{21}^0 = 227 \pm 16$ MHz, a metastable state with $R_{31}^0 = 594 \pm 221$ Hz, and an inter-system crossing (ISC) rate of $R_{23} = 176 \pm 21$ kHz, resembling those for SQEs in 2D h-BN. [3,34,40] Interestingly, for some emitters with similar spectral PL peaks, the excited state lifetime was much longer. Such an emitter can have a $\sim 5$ times longer excited state lifetime (Fig. 1e – black curve). We show below that this can be attributed to material dimensions smaller than our excitation wavelength. Emitters were completely stable under 594 nm excitation for $\sim 3$ hours of excitation without observable bleaching. Our suspension measurements therefore show that single QEs in BNNTs fabricated using the HTP method are associated with the material itself, discarding contribution from the substrate. Nevertheless, when switching to 532 nm excitation (focused on the same emitter), additional spectral features appear on the PL-spectra and the isolated emitter spectral features are absorbed in broader features (Fig. 1d - green curve). This is confirmed by the higher value of g²(0) ($0.61 \pm 0.01 > 0.5$), indicating more than one emitter (Fig. 1d). A structural defect, extending a line defect, as observed in TEM and cathodoluminescence measurements with different excited state energies (as depicted in Fig. 1c – inset), [17,25] can possibly explain this observation. Alternatively, due to the high density of compressed material in the micro-bundle and BNNT's large surface area, another emitter can be present inside our confocal volume. Using 532 nm excitation we were not able to isolate single quantum emitters. Furthermore, isolating single nanotubes with precision cannot be achieved using this method.

We therefore apply two methods for dispersion: First, using *N,N′*-Dimethylacetamide (DMAc), shown not to affect nanotube structure [44] (see Methods for more details). Second, we once again take advantage of the electric polarity of BNNTs – adopting a method of using oxygen plasmas from carbon nanotube research. [66] Namely, oxygen plasma, which slightly charges the surface, is used to treat the surface on which we rub the BNNT material afterwards (See Methods), thus circumventing solvent contaminants and allowing optical isolation of QEs. Using both methods, with 532 nm excitation alone, PL features of a single emitter could be isolated.. In Figure. 1f – curve 0, a representative PL of emitters which were measured using both these methods is displayed. However, emitter stability was affected using DMAc, limiting excitation time to ~ 1 hour before bleaching, hinting to possible site-specific modification as a result of the use of DMAc. [44] Nevertheless, BNNTs chemisorption of oxygen is not likely to take place using our sample preparation conditions at non-defective sites. However, defective sites such as Stone-Wales defects [25,26,27] are more chemically reactive due to local strain and bond frustration. [12,26] Previous research has also shown that QEs brightness in 2D h-BN is strongly affected by the supporting substrate (or lack of) [40], possibly influencing radiative rates and also stability. This leads us to believe that stability was affected by DMAc or by the substrate. So as to avoid artifacts, we repeated this procedure using only DMAc *without BNNTs* on $SiO_2$. No QEs could be found using DMAc alone. To further explore the role of oxygen (evident from our Raman/EDX measurements), we also plot the PL spectrum of a SQE (Fig. 1f – curve 1), created through oxidizing chemical etching of 2D h-BN (See Ref. 3). In addition, we plot PL measurements from two SQEs in v-$B_2O_3$ [58,59] (Fig. 1f – curves 2-3, see Methods for details). The resemblance of the emission wavelength is apparent, with a PSB detuning of 149 meV (1203 cm$^{-1}$) for etched h-BN and 155 meV/113 meV (1258 cm$^{-1}$/914 cm$^{-1}$, curves 2/3, respectively) for v-$B_2O_3$, although with a shift in the maximum peak wavelength. The h-BN/v-$B_2O_3$ detuning resemble best the hexagon ring in-plane $E_{2g}$ [3]/ E' [67] modes, respectively. A third peak is cut off by our 550 LP filter for curve 2. The v-$B_2O_3$ SQEs PL spectra (curves 2-3) are also similar to those reported for monolayer h-BN. [33]

Next, we analyze the PSB detuning from the ZPL for hybrid BNNT, which can be shifted by the supporting substrate [68] (10 meV for 2D h-BN) and by different boron isotopes. [20] Hence, we make a distinction between suspended and dispersed detuning. Comparing the PSB detuning from the ZPL between suspended and dispersed material reveals a different behavior: When dispersed on the substrate, the detuning is shifted up compared to the suspended wavenumber (energy). This is plotted in Fig. 1g, where the average wavenumber for suspended material is 821 cm$^{-1}$ (102 meV) and on $SiO_2$ is 1153 cm$^{-1}$ (143 meV). This shift behavior was consistent for all dispersion methods used in this manuscript and was independent of excitation wavelength (see SI for on substrate example using 594 nm excitation). As will be shown, the PL features (Fig. 1c, Fig. 1f) are related to hybrid material attached to BNNT. This shift hints to a strong impact the local environment has on the emitter. Interestingly, our detuning in the suspended/dispersed (on substrate) case, 821 cm$^{-1}$/1206 cm$^{-1}$ (102 meV/149 meV), is similar to the detuning seen for v-$B_2O_3$ emitters 914 cm$^{-1}$/1258 cm$^{-1}$ (113 meV/155 meV) curves 3/2 , respectively. Coincidently, the vibration of boron oxide anions ($^{10}BO_2^-$ and $^{10}BO^-$) which due to their bond strength, retain their integrity in complex chemical environments, have an excited state vibrational difference of 340 cm$^{-1}$ (43 meV) [69] similar to the wavenumber difference between our suspended and non-suspended material of 331 cm$^{-1}$ (41 meV). Alternatively, we can classify the suspended phonon modes to 2D h-BN $TO_\perp$ (ZO, along the c-axis) modes (see SI), possibly attenuated upon contact with the surface, making the $E_{2g}$ modes more prominent. Therefore, upon dispersion, a chemical change can occur in the emitter or the vibrational freedom is diminished (a schematic is depicted in Fig. 2g.), consistent with our observation of hybrid material on the tube exterior exposed to the environment, as will be shown. This behavior has not been observed before for QEs in suspended h-BN, [3,40] probably because of the relative small area the h-BN flakes were suspended on (on the scale of 10 microns) and the less exposed environment in a 2D h-BN lattice or possibly because of the different origin of emitters in this manuscript. In our experiment, the material is suspended over a scale of millimeters from the supporting aluminum needles. We demonstrate this behavior further below for suspended material with a supporting structure on a smaller scale.

### *SEM resolution of entwined BNNT SQEs*

We now turn to spatial localization and classification of QEs in the BNNT using SEM resolution. Using 532 nm excitation our spatial resolution is diffraction-limited at best to 200 nm, which makes distinguishing single nanotubes from nanotube bundles impossible. To overcome this, we combine higher resolution imaging of the SEM and optical laser excitation. To this end, we deposited 200 nm of indium tin oxide (ITO) on $SiO_2$, rendering the substrates transparent and conductive, thus making them compatible with both types of spectroscopy. In addition, quantum dot research has shown ITO to suppress emission blinking due to facile electron transfer (due to ITO's Fermi level) to possible trap states, thus blocking emitter electron transfer to these states once in the excited state. [70] Due to the proximity of our nano-material to the environment, we expect a similar effect. To probe suspended material, we also used a conductive gold TEM grid. The DMAc drop-casting procedure was applied to two substrate types: $SiO_2$ with ITO and a gold TEM grid.

Figure 2 depicts data for QEs originating from entwined BNNT material. Column (a) contains spectral and autocorrelation data, column (b) confocal optical excitation scans and column (c) SEM scans, where each enumerated line connects between the column features. Lines 1 and 2 depict data for suspended BNNT on a TEM grid and line 3 for BNNT on ITO. Surprisingly, on straight single nanotubes (without hybrid material) on the scale size of single nanotubes no QE emission could be exhibited. A comparison between line 1 and line 2 of Fig. 2 reveals that emission is *not* originating from single tubes suspended on the grid, but rather from the entwined BNNT. A closer look (Fig. 2c1/c2 – insets) shows that these tubes are a combination of a few tubes wrapped on what is most likely one. An autocorrelation measurement (inset Fig. 2a1) reveals that this is a single QE ($g^2(0) = 0.4 \pm 0.02 < 0.5$). On numerous instances, when single tubes were isolated, we were not able to see an emission signature associated with QEs. Similar ensemble features can be seen for BNNT on ITO (line 3 of Fig. 2). However, a cluster of curved nanotubes smaller than 1 micron contributes to emission. Therefore, nanotube termination, known to harbor defects, [26,27] can perhaps also play a role in emission. For Fig. 2a1, using an asymmetric peak fitting function (see SI) we subtract the broad shoulder of the PL spectra to identify six distinct peaks superimposed on the shoulder. An analysis (see SI) leads us to conclude that most probably the ZPL is at the range of 589 nm. In Fig. 2d we show a representative metastable decay behavior for this class of emitters for a similar SQE dispersed on ITO using the plasma dispersion method. Photo-dynamics reveal at least two components affecting the metastable decay ($|t| > \sim 40$ ns), one only prominent at high excitation powers (> 500 µW). This is plotted in Fig. 3d on a time logarithmic scale. We attribute the short component to a possible dark state, indicating an intermittent blink to trap states or a different emitter charge state, as seen for the silicon vacancy defect in diamond. [71] We get an excited state relaxation rate of $R_{21}^0 = 570 \pm 76$ MHz, a metastable state with $R_{31}^0 = 0.24 \pm 0.03$ kHz, and an ISC rate of $R_{23} = 140 \pm 8$ kHz. Photobleaching *was not observed* for this SQE, attributed to the stabilizing effect of the ITO substrate. Interestingly, two separate SQEs of this type show different photon emission counts power saturation behaviors (Fig. 3e). This can be attributed to different dielectric environments around the SQEs [3] or to variations in excitation/collection efficiency due to the circular quasi 1D nature of BNNT on which the emitter can be spatially located.

### *SEM resolution of hybrid material SQEs*

In Fig. 3 we analyze the spatial localization of emission from hybrid material dispersed using DMAc (see Methods). For lines (1) to (4), isolated nano-tubes clusters from a diluted BNNT DMAc solution, containing isolated segments of single nanotubes were probed. Line 5 is for a suspended cluster on a TEM grid. Columns (a) and (b) are similar to those is Fig. 2, whereas columns (c),(d) display SEM imaging with different resolutions, tilting angles and column (e) with backscattering imagery. Purple lines/circles connect between the column features, whereas cyan lines/rectangles mark entwined tube areas. From columns (b) and (c) we deduce that emission originates from the tube structure and not from the ITO substrate. We see a clear tendency of entwined twisted tubes (cyan rectangles) to exhibit concentrated emission areas, whereas isolated single nanotubes do not emit, consistent with our previous observations. All cyan rectangle areas resulted in a broad PL spectrum and not with the sharp narrow PL features displayed in column (a). The single QE in line 1 of Fig. 3 ($g^2(0) < 0.5$), reveals that emission originated from hybrid material and *not* from the continuous nanotubes. The brighter color of the hybrid material compared to the darker color of the tubes in the SEM backscatter images (Fig. 3e1, Fig. 3e2) can indicate a different crystallographic orientation, possibly due to a differing orientation of 2D h-BN, [13] causing the SEM electrons to scatter differently, [72,73,74] or different conductivity due to a changed ratio of less boron atoms in this area or a higher atomic number, like oxygen, [26,44] as indicated from our bulk Raman/EDX measurements (Fig. 1a). Hybrid nanomaterial comprising of 1D BNNT and 2D BN oxide platelets have also been reported for high temperature annealed BNNT. [12] The minute size of the hybrid material is insufficient for EDX spectroscopy. The good dispersion, tube structure isolation and the long magnetic stirring process involved in dispersion (See Methods) [44] evident from Fig. 3 leads us to believe that the hybrid material is *part* of the tube structure and not attached via weak non-covalent bonds, normally holding non-defective BNNT walls together. These are also similar in appearance to hybrid material identified from the same vendor using TEM, confirming attachment between the BNNT and the hybrid material. [13] A similar analysis is valid for the purple lines 2-4 of Fig. 3. The point-like emission originates from hybrid material smaller than our excitation wavelength (Fig. 3e2). Therefore, the dielectric nano-antenna effect can also govern the emission mechanism for this QE class, affecting emission rates and saturation behavior, [75] thus explaining our diverse lifetimes in Fig. 1e. The PL spectrum in Fig. 3a(1,4,5) are consistent with that of Fig. 1f (curves 0/1) and structurally with that in Fig. 1e (orange curve) and for Fig. 3a(2,3) to those seen in suspended material where the PSB was absorbed in the ZPL (not shown). Tilting the SEM image by 30 degrees (Fig. 3d1 – bottom inset), reveals the entwined tube areas in more detail and the hybrid material. Interestingly, line 5 in Fig. 3 shows that suspended entwined BNNT tubes can seemingly also give sharp spectra. Nevertheless, this is the only instance where we have seen this behavior, leading us to believe that hybrid material is present in the suspended micro-tube. The PSB detuning from the ZPL is 1322 cm$^{-1}$ (164 meV), which we categorized in the ranges of 'On substrate' in Fig. 2h. However, in line 5 of Fig. 3 the BNNT material is located at a distance of ~ 1 μm from the supporting frame of the gold TEM grid and from densely entwined tubes (not shown). Therefore, the PSB detuning behaves like the 'On Substrate' case.

*Artificial curving of 2D h-BN*

It has been demonstrated that SQEs in h-BN can be spatially correlated with wrinkles on an h-BN flake. [39] Therefore, to show that BN nano-material artificial curvature can be spatially correlated with QE location, we fabricated sharp diamond pillars [76] 554 nm high with the top/bottom base diameters of 93 nm/200 nm, respectively. Figure 4b displays a SEM scan of a typical pillar. Single crystal 2D h-BN flakes [77] with different thicknesses were transferred on top of the diamond substrate so that they would artificially curve on the pillars and then annealed (see SI for details). Some pillars also pierced the h-BN membrane. Figure 4a/4c shows a reflection (no wavelength filters) / 550 LP filtered 532 CW excitation confocal scan, respectively. To gain higher structural resolution insight, in Fig. 4d/4f, a SEM scan of this flake, 30 degree tilt/top view, respectively, is presented. Pillars in which QEs were observed are marked with numbers between '1' to '6'. In addition, QEs could be found in *perimeters* of flakes. Comparing between Fig. 4a/4c/4d/4f, QE '1' is spatially localized with the curving of the h-BN membrane on the pillar and, as opposed to Ref. 3, no *perimeters* can be seen. Both QE '1' and '6' are in a non-pierced area and exhibit a 'tenting' effect, however, QE '6' is also on the h-BN edge (insets of Fig. 4d,f). An AFM scan (see SI) reveals the peak of QE '1' to be ~ 1 μm above the h-BN flake flat plane fitting exactly the convergence height from our pillar bases (1 μm), therefore the top is not in contact with the pillar (of 554 nm height), with an opening angle of 16.7 degrees, wider than the pillar bases convergence point angle of 11 degrees. Presumably, curving commenced when the flake started to touch the pillar bases on one side, which could have been modified afterwards by annealing. An autocorrelation measurement of QE '1' is displayed in Fig. 4d, proving its single nature ($g^2(0) =$

$0.37 \pm 0.02 < 0.5$). We note that due to the small diameter of the pillars, pillars without h-BN flakes emit background upon laser excitation. Autocorrelation measurements using the same power (50 µW) were also conducted on these pillars revealing no quantum signature - validating quantum emission association to the h-BN flakes. Pillars with h-BN on top near flake boundaries (~ 1 µm away) also showed QEs clustering which resulted in more than one emitter. This is seen for QE '6' (Fig. 4e), yielding $g^2(0) = 0.82 \pm 0.01 > 0.5$. QEs '2','3','4' and '5' are clearly pierced, (Fig. 4a,d,f). Nevertheless, these areas also displayed a quantum signature, consistent with previous work, where SQEs were measured in an area of a chemically etched hole or boundaries in the h-BN flake.[3] We discuss below in detail the difference between curving (tenting) and piercing. For pillars marked in black lines in Fig. 4d, no QEs were found. A rough comparison to the other pierced pillars (Fig. 4d) shows that they have not penetrated as deeply into the flake as the other pillars, inducing less curvature in the surrounding area, possibly hinting further to the role of curvature or other mechanisms in QE formation. To further demonstrate the role of *perimeters* and curvature abundant in the h-BN (or created 'naturally' by exfoliation), an area marked with 'Fold' in Fig. 4f is shown and its quantum nature (Fig. 4e – autocorrelation measurement). An AFM scan shows the fold height to be ~ 250 nm with a slope angle of ~ 2 degrees. Figure 4g shows the PL spectrum for QEs from pillars '1' to '6' (curves 1 to 6), and for the QE marked with 'Fold' (QE '8') and another QE from a different BNNT sample (QE '7'). Figure 4h displays the PSB detuning from the ZPL for all QEs '1' to '5' (omitting QE '6' from the analysis due to its ensemble nature). We identify the same two-peak structure as identified for hybrid material in BNNT when dispersed on the substrate (Fig. 3a1, a4 / Fig. 1g). In addition, a similar average wavenumber (energy) detuning is seen 154 / 149 meV (Fig. 4h / Fig. 1g – red line, respectively). Intuitively, a pillar which would cause the 2D h-BN flake to curve near its edge would have quasi-1D features, similar to the structure one finds in nanotubes. Comparing between the PL spectra of QEs from pillar '6' and QEs from a BNNT sample (Fig. 4g curves 6 and 7, respectively) we see similar spectral features, although the peak maxima are shifted with respect to one another. Similar local maxima are marked by arrows (up/down, BNNT/h-BN pillar, respectively). Therefore, curved areas can result in resembling PL spectra in h-BN and BNNT consistent with studies showing that optically BNNT can behave as curved 2D h-BN. [17] Furthermore, the PL spectral features are different from the QE in the area marked as 'Fold' (curve 8), emphasizing the difference between edge artificial curving and 'natural' curving in h-BN. We also attempted to see QEs on thick h-BN flakes (> hundreds of nm). On pillars on which thick h-BN was placed, QEs could not be seen (not shown). Due to the higher thickness we think that the flake is mechanically more resistant to deformation and piercing, and as such QEs could not form. Finally, h-BN flakes were exfoliated on a hemisphere of $ZrO_2$ to insure the flake contains long segments of curved areas. The effect of as-exfoliated compared to flakes exposed to a liquid environment was tested, confirming the sensitivity of BN processing chemicals (see SI).

**Discussion**

The classification of emission from **hybrid material** (Fig. 3) can have two interpretations: A simplistic one where the defect responsible is presumably contained only in the hybrid material. However, the formation of hybrid material attached to the nanotube structure is an indicator of nanotube reconstruction which would disrupt the nanotube structure due to defects, as previously observed. [73] Thus, possibly a vacancy defect, shown to cause $sp^3$ bonding between BNNT walls [18,73] during the growth phase of the tube, might be located on the exterior BNNT wall and thus $sp^3$ bond the whole complex. Electron energy-loss spectroscopy (EELS) spectroscopy has also shown that curvature in BNNTs can induce $sp^3$ hybridization. [78] The hybrid material could be 2D h-BN [13] or possibly boron oxide, [44,12] due to its different color in the SEM backscatter detector. The narrow PL features in the nano-tubes clusters in Fig. 3 seen where hybrid material is present, demonstrates it plays a vital role for this type of emitter formation. SQEs seen in v-$B_2O_3$ strengthens the role oxygen defects can play in boron systems, proving that similar PSB detuning can be obtained also in v-$B_2O_3$ to that of hybrid BNNT SQEs. Moreover, the planar structure of $B_3O_6$ and the similar PL to that of monolayer h-BN SQEs [33] hints to an analogy between them. The oxygen atomic mass would not compromise significantly the 2D h-BN host lattice mass, yielding similar Raman signatures. [3] The abundance of SQEs and the larger surface area exposure to the environment of BNNT material compared to a bulk h-BN crystal also strengthen these observations. Possibly, nano-$B_2O_3$ areas present in the BN lattice can be partially responsible for emission in previous 2D research attributed to h-BN.[3,33,40,36,34] Interestingly, $BN_2O$ and B-$NO_2$ point defects retain the h-BN's planar structure while B-$O_3$ has a random orientation, [56] possibly resolving recent optical polarization measurements. [3,40] Another possibility would be BN di-vacancies, [39] reconstructing to Stone-Wales [24] defects [3,33,27] (seen in TEM, BNNT [25,26]/ h-BN [79]), point defects or combinations with oxygen. [12]

For emission from **BNNT entwined material** (Fig. 2), oxygen structures calculations ($B_3O_6$ structures, present in v-$B_2O_3$ [58,59]) yielded energy states in the visible range (~ 2 eV). [21] We can also interpret interlayer bonding, due to the inability to see SQEs on single straight tubes lacking interlayer bonding or curvature, as a source of emission.

At a first glance the structural effect of piercing and **curving of h-BN on pillars** (Fig. 4) are different, however they can have similar effects. Namely, the former creates a micro-defect/rupture in the membrane visible using SEM microscopy and the latter lowers the formation energy of defects, thereby making a point atomic defect formation more likely, visible using TEM microscopy. Full piercing would create curvature in the membrane, surrounding the pillar (as shown in Fig. 4c), thereby making it also a combination of both effects. Nevertheless, the natural abundance of point defects in h-BN [77] or absorbed atoms (as demonstrated in the SI), [49] with the combination of curvature may give rise to QEs. Possibly, oxygen healing or interaction with other functional groups, [80] stabilized by annealing, occurs on the curvature induced or pierced (*perimeter*) defective sites which are very reactive, [81] causing oxidation to be energetically favorable, [82] even more so favorable with the abundance of surface boron vacancies causing atomic curvature. [83] Alternatively, similar to the entwined BNNT case, interlayer bonding between h-BN layers [83,3] and strain modified energy states can also play a role.

**Conclusions**

We uncover a new material of the BN family for SQE use. Our results strengthen previous observations in 2D h-BN, [3] correlating *perimeters* and the role of thinner material with the increase of QEs, due to the lower energy formation of defects on the exterior. Isolating SQEs was possible upon the reduction of excitation energy and of BNNT material or their dispersion, demonstrated by three methods: BNNT 'string' pulling using 594 nm excitation or DMAc/oxygen plasma dispersion using 532 nm excitation. The high quality of fabricated BNNT / hybrid material, combined with dispersion using the oxygen plasma method, allows literally 'out-of-the-box' emitters avoiding the use of solvents and any further fabrication processing (such as annealing) to work with SQEs. In addition, the combination of 1D material with hybrid material allows easy suspension, avoiding solvent and substrate contaminants, which could advance BN QE research further. The lack of processing contaminants is a key milestone for unraveling the atomic structure giving rise to QEs, with precision atomic spectroscopy technologies available. The minute nature of the material as seen in SEM, offers potential background-free QEs upon suspension. We find that conducting surfaces, such as ITO, are advantageous in avoiding QE excited electrons falling into possible trap states. For the first time, we show a one-to-one optical correlation of visible wavelength SQEs in BN with SEM spatial resolution, for suspended and on-substrate material, possibly confirming previous observations in h-BN. [33,34,3,35,37,38,40,41,42] Material with features smaller than our excitation wavelength elucidates dielectric nano-antenna effect, varying excitation lifetimes. [75] The 1D nature of BNNT can allow bio-adsorption of nanotubes with QEs for intra-cell imaging, making a single BNNT tube a natural QE carrier, using less energetic laser excitation. The spatial localizing of curved areas and QEs puts h-BN under a similar umbrella with QE observations seen in other semi-conducting TMDC QEs [84,85] (with the distinction of the large bandgap and 2D h-BN annealing procedures). This also allows us to interpret previous observations of creation of QEs in 2D h-BN: induction by electron [34,40]/ion [3]/FIB [42] irradiation, chemical etching and spatial localization on edges and between different emitter layers. [3] Namely, all of these methods induce curvature/strain and the spatial localizations are correlated with areas (such as edges) that are more likely to be curved than perfectly flat, and thus would be more reactive. We can also think of annealing, typically used to stabilize QEs, as a mechanism to 'fixate' the curvature of bubbles or to concentrate large bubbles near *perimeters* in a 2D flake, thus localizing smaller bubbles inside the flake [86] and thus resulting in more localized QEs. We prove that h-BN PL features are sensitive to the external environment, such as solvents/liquids (see SI), and therefore some of the SQEs in commercial BN dispersed in liquids [33,41] could originate from the interaction of the liquids with surface vacancy defects. Bulk BNNT can be viewed as a 'cheese' with holes, namely, the environment surrounds the whole bulk. Further research on techniques to manipulate SQEs energy levels in BNNT could lead to sensing routes of these emitters, exposed to the environment. Isotopic engineering of the boron species can be done to dramatically reduce

SQE phonon-electron broadening.[20] Our results indicate that artificially bending BNNTs might be beneficial for engineering QE formation in BNNTs, but this is beyond the scope of this work. These results further highlight the high quality of material which can be obtained with high temperature fabrication for QEs in BN systems. Implementation of the techniques in this work with TEM and STEM technology is straightforward, highlighting the potential of low-dimensional wide bandgap semiconductors as emerging platforms for deep bandgap point-defect engineering.

**Methods**

*Bulk BNNT sample preparation and measurement*

SiO$_2$ slides were thoroughly cleaned in a peroxymonosulfuric acid (H$_2$O$_2$:H$_2$SO$_4$) solution for 2 hours and afterwards rinsed in DI water, isopropyl alcohol and dried in pure N$_2$ gas. Glue was placed on the cover slides and a substantial amount of BNNT was extracted from the bulk "cotton"-ball, and placed on the glue. The sample was mounted to a room temperature setup at ambient conditions. We avoided excitation of the area the BNNT was glued to the cover slide and concentrated on areas tens of microns away from the surface. Photo emission was detected using two avalanche photo diodes (APDs) in a Hanbury Brown and Twiss configuration.

*BNNT suspension using needles*

A smooth steel board with an elongated line drilled in the center with two separate square SiO$_2$ slides with dimensions of 0.5 x 0.5 x 0.5 cm glued on top was prepared. The two square slides had a ~ 1 mm spacing between them and were placed parallel to each other on the opposite edges of the x axis of the board substrate so that on the y axis there is a non-intersecting area above the drilled lined. Afterwards two conductive needles were glued with their bases glued to the x axis of each of the SiO$_2$ slides. From bulk BNNT a small string-like amount was isolated with diameters on the micron scale using tweezers. Each end of the BNNT string was placed on the conductive needles. After contact with the conducting needles no further fixation of the BNNT string was needed.

*BNNT QE Dispersion using DMAc*

For dispersed material in Fig. 1 and Fig. 2, the exact same method of BNNT dispersion was used as in Ref. 44. Magnetic stirring was applied via a Teflon-coated magnet for 4 days. For Fig. 3 we used a substantially reduced amount of BNNT, namely the smallest micro-bundle one can extract simply using a tweezer. The same DMAc stirring procedure was afterwards applied. For dispersion on $SiO_2$ and $SiO_2$ + ITO, $SiO_2$ was first cleaned in a peroxymonosulfuric acid ($H_2O_2$:$H_2SO_4$) solution for 2 hours similar to the procedure described previously. In order not to damage the ITO, $SiO_2$ + ITO were cleaned in hot acetone at 80 °C for 1 hour, and hot isopropyl alcohol at 65 °C for 1 hour, finally rinsed in cold isopropyl alcohol and dried in pure $N_2$ gas. The TEM grids were placed in a holder where the TEM grid is suspended on the grid areas. All substrates were placed on a hot plate and preheated to temperatures between 185 °C to 200 °C. The DMAc + BNNT solution was always pre-sonicated for 30 min. (in accordance with Ref. 44) before drop casting the solution. Drop casting was done in a fume hood and the hot plate was left on for 5 minutes. Afterwards, the hot plate was turned off and let to cool to room temperature. We could re-use dispersed DMAc + BNNT solutions weeks after dispersion, since BNNT did not create sediments at the bottom of the solution, in accordance with Ref. 44.

*BNNT QE Dispersion using oxygen plasma*

$SiO_2$ and $SiO_2$ + ITO slides were pre-cleaned using the procedures described above. The samples were then exposed to $O_2$ plasma for 5 minutes in vacuum. This further cleans the surface from undesired organic contaminants (which can obstruct optical excitation) and slightly charges the substrate surface, increasing BNNT affinity to the substrate. Therefore, by simply rubbing a small portion of material, BNNT is dispersed on the surface to allow optical isolation of QEs. Immediately after exposure, a sample amount of BNNT was extracted from the bulk using a clean tweezer, and the side opposing the tweezer contact point was rubbed on the substrate while viewed under the microscope to see the BNNT placement on the substrate. The sample was then placed on an unheated hot plate (to promote adhesion of material to the substrate) which was then heated to 70 ⁰C. Once 70 ⁰C was reached, after 5 minutes we turned off the hot plate and let it cool to room temperature.

*BNNT EDX*

EDX was performed using an SDD detector with a NSS system from Thermo Fisher.

*v-$B_2O_3$ preparation and measurements*

The material was acquired from the vendor 'Suprapur'. $SiO_2$ cleaned in peroxymonosulfuric acid as described previously was used as a substrate. Glue was placed on the substrate and immediately afterwards bulk v-$B_2O_3$ (~ cm sized) was placed on top. After waiting for the material to fixate we probed v-$B_2O_3$ using the 532 nm laser and a 550 LP filter far away from the substrate contact point. Point like emitters could be immediately seen once probed. To verify that we were looking inside the bulk material once focused on an emitter and not on the surface we reduced the excitation power to ~ nw and removed the filter, therefore probing the reflection. This confirmed we were focused in the bulk of the material. Afterwards PL spectra was collected and anti-bunching measurements were performed.

**Corresponding Author**

*Inquiries should be sent to the following e-mail address: a.finkler@physik.uni-stuttgart.de

**Acknowledgements**

The authors acknowledge support from the Max Planck Society, the DFG and the EU via the project DIADEMS. AF and YK acknowledge financial support from the Alexander von Humboldt Foundation. JHS acknowledges financial support from the EU graphene flagship. K.W. and T.T. acknowledge support from the Elemental Strategy Initiative conducted by the MEXT, Japan and JSPS KAKENHI Grant Numbers JP26248061, JP15K21722 and JP25106006. We thank Torsten Rendler, Matthias Widmann and Nabeel Aslam for technical assistance and Rainer Stöhr for fabrication assistance. We also thank Felipe Fávaro de Oliveira and Seyed Ali Momenzadeh fruitful discussions. We would like to express our gratitude to Felicitas Predel from the Stuttgart Center for Electron Microscopy (StEM) for performing all EDX measurements. We are also grateful to Audrius Alkauskas for a critical reading of the manuscript and fruitful discussions.

**Author Contributions**

N.C., A.F and D.D. conceived and designed the experiments. N.C. and A.Z. prepared the BNNT samples. N.C. performed all confocal and SEM measurements and together with A.F. and D.D. analyzed the data. A.F. fabricated the diamond pillars, Y.K. performed the Raman measurements, h-BN pillar sample preparation and annealing. B.S. grew the ITO films. T.T. and K.W. grew the single crystal h-BN material. J.W. and J.H.S. supervised the project. N.C. performed the experiments and wrote the manuscript with input from A.F., D.D., J.H.S. and J.W.

**Competing Interests**

The authors declare that they have no competing interests.

**Data Availability**

The datasets generated during and/or analyzed during the current study are available from the corresponding author on reasonable request.

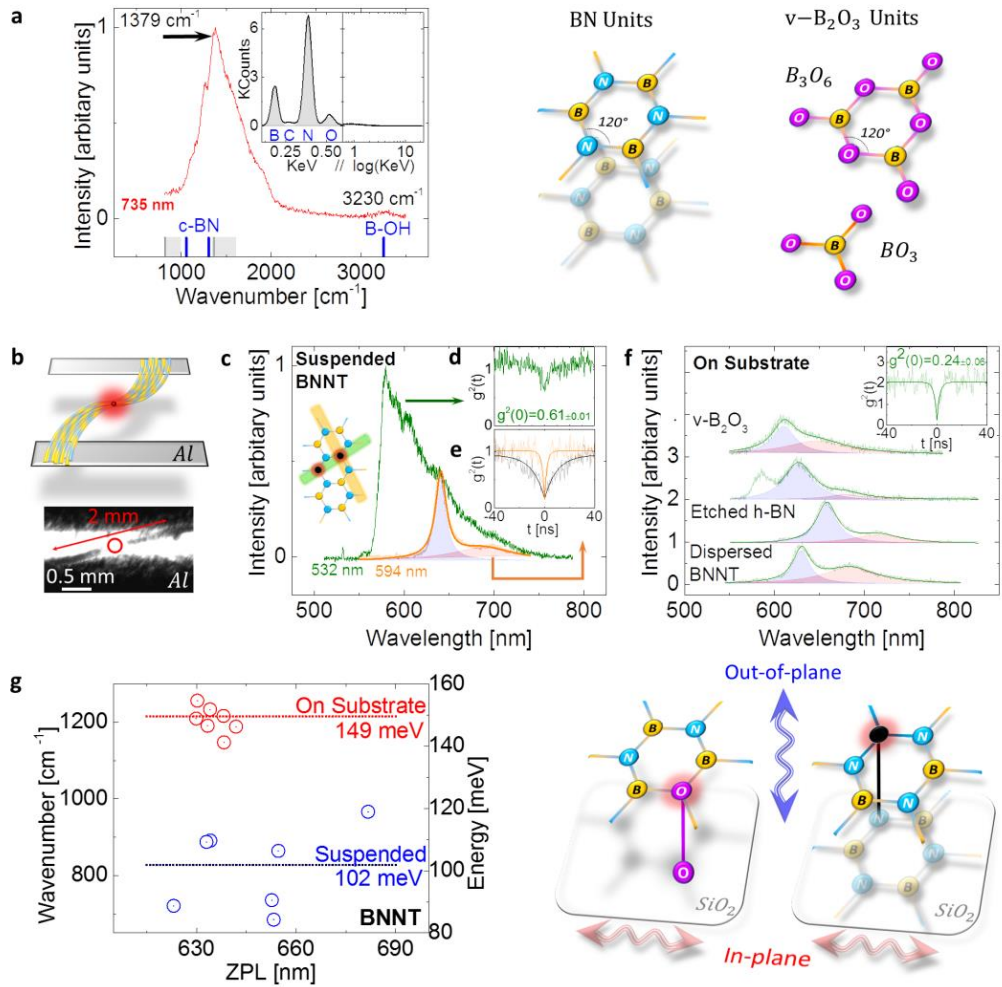

*Figure 1: (a) Raman-Stokes shift measurement using 735 nm excitation for bulk BNNT (red curve). Known Raman shifts are displayed for clarity for h-BN and BNNT (overlapping grey squares and grey lines), c-BN (adjacent blue lines) and B-OH (blue line). Insets display autocorrelation measurements. (1) and (2) are for BNNT dispersed with DMAc on a TEM grid (2) and (3) BNNT dispersed with DMAc on ITO. (c1 - inset) graphically enlarged image color coded in cyan for clarity showing a few nanotubes wrapped on a single nanotube, covering the TEM grid. (c2 – inset) a similar singular straight tube (d) $g^2(t)$ metastable state decay curves for 532 nm CW excitation in logarithmic scale. Blue curve represents low power 40 µW excitation, red curve high power 500 µW excitation. A second decay component is visible on the high power curve. (e) kilo counts per second for two SQEs with the same PL spectra structure, both display different counts as a function of power.*

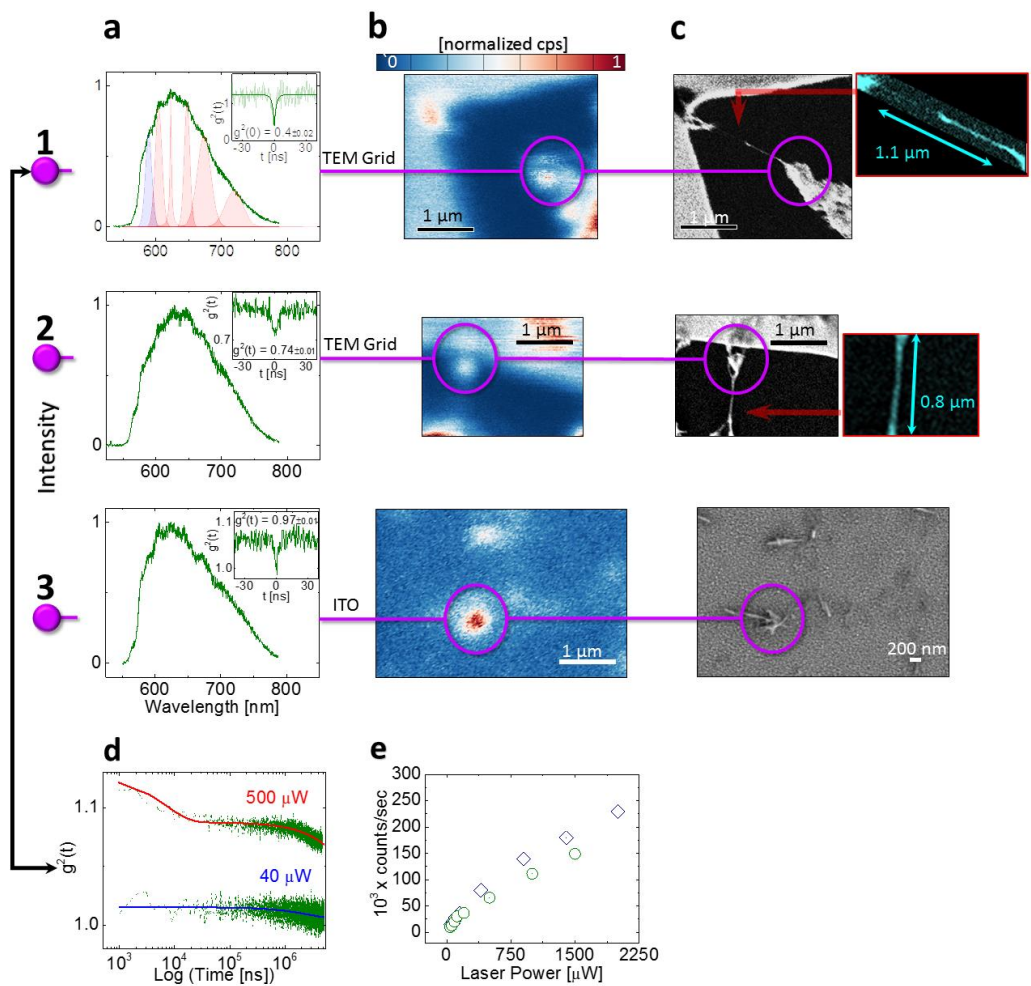

*Figure 2: (a to c)(1 to 3) compare the PL spectra (a), optical confocal scan (b) and SEM resolution images (c), correlating QE spatial features and emission properties. Substrate types are denoted above each line. Column (a) (1 to 3) displays PL spectra using 532 nm CW excitation. Insets display autocorrelation measurements. (1) and (2) are for BNNT dispersed with DMAc on a TEM grid (2) and (3) BNNT dispersed with DMAc on ITO. (c1 - inset) graphically enlarged image color coded in cyan for clarity showing a few nanotubes wrapped on a single nanotube, covering the TEM grid. (c2 – inset) a similar singular straight tube (d) $g^2(t)$ metastable state decay curves for 532 nm CW excitation in logarithmic scale. Blue curve represents low power 40 µW excitation, red curve high power 500 µW excitation. A second decay component is visible on the high power curve. (e) kilo counts per second for two SQEs with the same PL spectra structure, both display different counts as a function of power.*

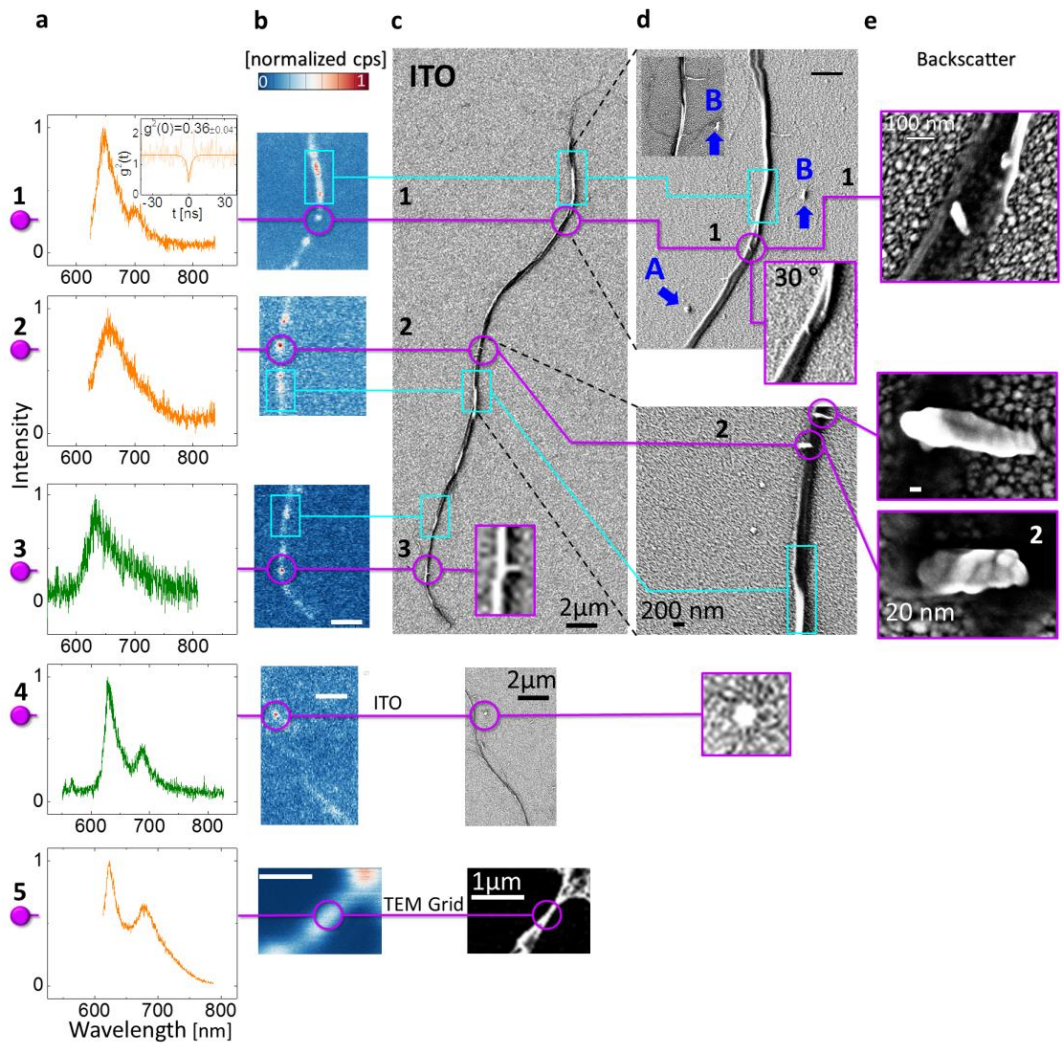

*Figure 3: (a to e)(1 to 5) compare the PL spectra (a), optical confocal scans (b) and SEM resolution images (c)(d)(e), correlating QE spatial features and emission properties for BNNT dispersed using DMAc on SiO2 with 200 nm ITO (1-4) and on a TEM grid (5). (a, ,b) (1 to 2) displays PL spectra using 594 nm CW excitation , (a)(1) Inset displays an autocorrelation measurement. (a,b)(3 to 4) PL spectra using 532 nm CW excitation (c) secondary electron SEM image of a nano-tube cluster, some containing segments of singular nanotubes. (d) Higher resolution SEM image of segments, revealing single nanotubes branching from the nanotube cluster (upper inset) and twisted tubes when tilting the detection by 30 degrees (bottom inset). Note – d4 was graphically enlarged. "A" and "B" (in blue) are markers for the eye. (e) SEM backscattering images revealing hybrid material attached to nanotube clusters. ITO grains are visible on the substrate. (b,c)(1 to 4) have the scale of 2 µm, (b,c)(5) have the scale of 1 µm.*

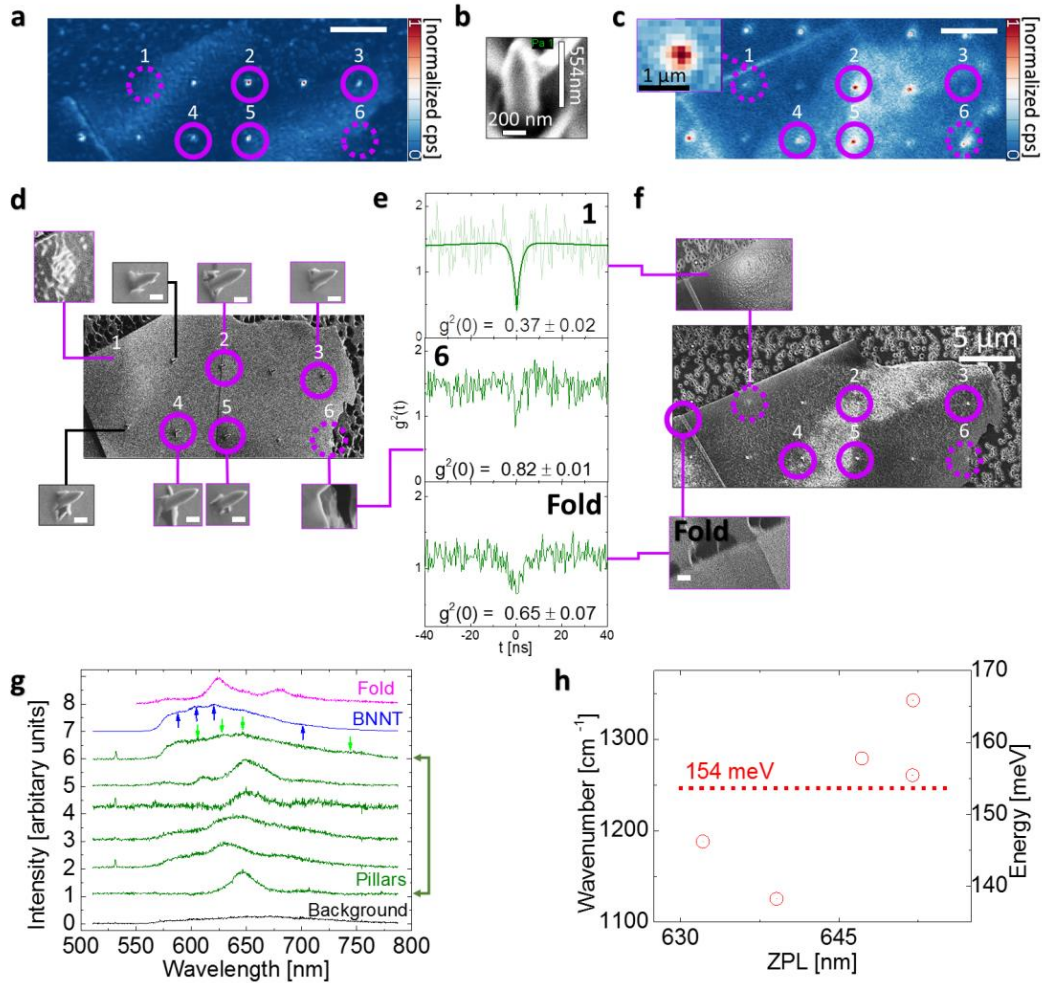

*Figure 4: (a) 532 nm CW reflection measurement of a h-BN flake placed on top of diamond pillars. Pillars where QE were observed are marked with purple circles from '1' to '6'. Pierced pillar areas are clearly seen. (b) SEM image of a typical pillar (c) 550LP filtered area of (a). (a),(b) and (c) have the same scale of 5 μm. (d) 30 degree tilted SEM image of the h-BN flake, scale bar marked with a white line is 200 nm long. The inset of pillar '1' was graphically filtered highlighting the 'tenting' topography of the flake. Piercing is seen on other pillars. For pillar '6' (inset), 'tenting' is seen at the edge of the flake (a dim gray half circle is seen). (e) Autocorrelation measurements from pillars '1', '6', and a folded area of the flake. (f) Top SEM view of the h-BN flake, pillar '1' – inset, graphically filtered image highlighting 'tenting' effect. Fold inset, magnification of folded region. (g) PL spectra for QE, '0' is the diamond pillar background where h-BN does not cover the pillar (normalized relative to QE PL spectra), '1' to '6' are the PL spectra of QEs, '7' is a QE PL spectra from a different BNNT sample for which background was subtracted. Arrows indicate resembling peaks (h) Detuning of the PSB from the main peak for QEs '1' to '5'. Left scale is in wavenumbers, whereas the right scale is in energy.*